\documentclass[twocolumn,showpacs,preprintnumbers,amsmath,amssymb]{revtex4}

\usepackage{epsfig,psfrag}
\usepackage{dcolumn}
\usepackage{bm}
\usepackage{graphicx}
\usepackage{color}
\usepackage{dsfont}

\begin{document}
\title{Fermi liquid theory for SU(N) Kondo model}
\author{Christophe Mora}
\affiliation{Laboratoire Pierre Aigrain, ENS, Universit\'e Denis Diderot 7, CNRS 
 24 rue Lhomond, 75005 Paris, France}

\begin{abstract}
We extend the Fermi liquid theory of Nozi\`eres by introducing the next-to-leading order 
corrections to the Fermi liquid fixed point. For a general SU(N) Kondo impurity away from
half-filling, this extension is necessary to compute
observables (resistivity, current or noise) at low energy.
Three additional  contributions are identified
and their coupling constants are related using an original (and more complete)
 formulation of 
the Kondo resonance {\it floating}. In the conformal field theory language, a single
cubic operator is proposed that produces the same three contributions with the same
coupling constants. Comparison with an exact free energy expansion further relates the
leading and next-to-leading order corrections so that a single energy scale,
the Kondo temperature,
eventually governs the low energy regime. We compare our results at large $N$ with
the approach of Read and Newns and find analytical agreement.
\end{abstract}


\maketitle

\section{Introduction}


The fascination exerted by the Kondo model~\cite{hewson1993} is probably due to the large variety of 
theoretical techniques invented to describe it. In fact, it has proven quite difficult 
to find a single approach that alone explains all features of the Kondo model.
This is particularly true in out-of-equilibrium situations~\cite{oguri2001+meir2002+rosch2003+kehrein2005+mehta2006+anders2008}, 
for example when a bias voltage is applied to a source-drain  setup. 
The seminal papers of Nozi\`eres~\cite{nozieres1974a,nozieres1974b+nozieres1978} 
on the Fermi liquid (FL) theory have provided a
remarkable insight into the low energy regime of the Kondo model. Based on a phenomenological
picture,  this approach contains nevertheless all relevant physics and leads to predictions
that are exact, albeit perturbative. The most famous example is certainly the Wilson ratio,
predicted by Nozi\`eres~\cite{nozieres1974a} 
 to be exactly two in agreement with numerical estimates
by Wilson~\cite{wilson1975}. Finally, the FL picture provides a straightforward tool to study analytically
the out-of-equilibrium regimes.

The FL approach has been recast later in the more formal language of conformal field
theory (CFT) by Affleck and Ludwig~\cite{affleck1990,affleck1991,affleck1993}. 
In this framework, the quasiparticles of the FL constitute
a boundary free field theory~\cite{affleck1990} 
 which is  the  infrared strong coupling fixed point of the Kondo model.
The low temperature regime is then dominated by the leading irrelevant operator at this fixed
point and the results~\cite{affleck1993} are in  complete agreement with Nozi\`eres.
More recently, elaborating on a more involved
version of the Bethe ansatz, Lesage and Saleur~\cite{lesage1999a+lesage1999b} 
were able to justify the FL theory for ordinary
SU(2) Kondo and to extend it to all leading irrelevant operators.
To be more exhaustive, we shall mention the work of Yosida and Yamada 
published independently from Nozi\`eres. In a
series of papers~\cite{yosida1970+yamada1975a+yosida1975+yamada1975b} 
on the parent Anderson model, they did a thorough analysis 
of perturbation theory in the interaction term $U$. They derived general
low energy properties for the self-energy
that proved the Fermi liquid picture extending it to finite $U$.
The extension of their work to the second order low energy corrections is
yet an unsolved problem.
Aside from these works and perhaps surprisingly, the FL theory as presented by Nozi\`eres 
was not pursued much further~\cite{hewson1993,hewson1994},
probably because no simple means were known  to relate  the different phenomenological 
coefficients of the theory. Following studies have started instead to focus on more exotic non-Fermi
liquid regimes~\cite{affleck1991,affleck1993,nozieres1980}.

Originally discussed for an ordinary spin-$1/2$ impurity with SU(2) symmetry, the FL fixed point
constitutes more generally the low energy limit of the Kondo model for a SU(N) hyperspin 
impurity. The value of $N$ tunes the relative importance of the different low energy processes.
 This SU(N) Kondo model is called the Coqblin-Schrieffer model~\cite{coqblin1969} 
for a single-electron impurity.
Both this model and its parent Anderson model have exact Bethe 
ansatz solutions~\cite{andrei1983+tsvelik1982+rasul1982,schlottmann1982+tsvelik1983}.
The SU(4) case has a particular experimental relevance with recent achievements in vertical
quantum dots~\cite{sasaki2004} and carbon 
nanotubes~\cite{herrero2005+makarovski2007a+makarovski2007b,delattre2009}.
In those experiments, an orbital degeneracy might combine with the usual spin-$1/2$
to form an intricate SU(4) symmetry. 

The conventional Fermi liquid description contains only the leading irrelevant  
 operators of dimension $3$, also  linear in $1/T_K$ where $T_K$ is the Kondo temperature.
These operators include a combination of an elastic channel with an inelastic one.
The ratio of elastic to inelastic scattering amplitudes is  
fixed by the Friedel sum rule~\cite{langreth1966} or more generally by the principle of
{\it floating} of the Kondo resonance that we shall detail in the core of this article.
This fixed ratio can also be shown to be a consequence of the vanishing charge susceptibility
on the dot~\cite{hewson1994}.
The conventional FL approach, as we described, is sufficient to compute observables that have a linear
energy ($k_B T$, $e V$ or $\mu_B B$) dependence, hence the success in the determination
of Wilson's ratio even for a general SU(N) symmetry~\cite{nozieres1980}.
However, for observables with a quadratic behavior such as the resistivity or the conductance,
the addition of dimension-$4$ operators becomes necessary.
The ordinary SU(2) Kondo effect is peculiar in this respect since the coefficients of
these new dimension-$4$ operators identically vanish.

The purpose of this paper is to extend the conventional FL approach by introducing
the full set of dimension-$4$ operators with their coefficients.
Within the theoretical framework proposed by Nozi\`eres, this second generation
adds three terms to the conduction electron phase shift. One represents elastic
scattering and two inelastic scattering involving the excitation of one and two
electron-hole pairs.
The ratios between the coefficients of the three FL corrections are then fixed by using
the {\it floating} of the Kondo resonance.
Let us emphasize that the picture built in this article for the Kondo resonance
{\it floating} extends the initial vision of Nozi\`eres. Not only the peak of the resonance
is tied to the Fermi singularity but also the {\it whole structure} of the resonance.
We also investigate how this translates into the CFT language. A single dimension-$4$
operator is identified with SU(N) invariance. Its expansion on the electron fields
recovers the aforementioned three processes with the same coefficient ratios.
Last step of the analysis, the free energy as a function of generalized magnetic
fields can be easily calculated within the FL theory including all dimension-$3$
and dimension-$4$ operator sets. Comparing the result with the exact solution
obtained from an alternative Bethe ansatz~\cite{bazhanov2003}, 
the ratio between dimension-$3$ and
dimension-$4$ corrections can be determined.
All coefficients are  finally related to each other so that, as expected, universality is recovered as
$T_K$ remains the only energy scale in the problem. This completes our full characterization
of the low energy FL theory for the Kondo SU(N) model.
We stress again that this work does not modify (and therefore does not contradict)
the ordinary SU(2) analysis~\cite{glazman2005} since the new FL corrections are vanishing in that case.
However these new corrections are fundamental in the more general SU(N) case
where particle-hole symmetry is broken.

The idea of introducing the next-to-leading order FL corrections was first formulated in Ref.~\cite{vitu2008},
although incompletely. It was however not taken into account in Ref.~\cite{lehur2007}.
The current and the noise through a SU(N) Kondo quantum dot were calculated in Refs.~\cite{vitu2008,mora2008},
with a correction in~\cite{mora2009} on the basis of this work.
The rest of this article is organized as follows: the new FL corrections are introduced
in the usual FL framework in Sec.~\ref{sec:FL} with an emphasis on the Kondo {\it floating};
and in the CFT language in Sec.~\ref{sec:CFT}. Sec.~\ref{sec:ba} compares the free energy with
the exact Bethe ansatz solution. Sec.~\ref{sec:1n} proceeds with a $1/N$ expansion which coincides
with the field theoretical large $N$ approach of Read and Newns~\cite{newns1987}.
Sec.~\ref{sec:c} concludes.

\section{Fermi liquid theory}\label{sec:FL}

Let us define the problem more precisely. The starting Kondo
Hamiltonian is (we follow Einstein convention for the capital superscripts)
\begin{equation}\label{kondo}
H = \sum_{k,\sigma=1 \ldots N} \varepsilon_k  b_{k\sigma}^\dagger b_{k\sigma}
+ J_K  S^A \, \sum_{k,k',\sigma,\sigma'} b_{k\sigma}^\dagger 
T^A_{\sigma,\sigma'} b_{k'\sigma'},
\end{equation}
with the dispersion $\varepsilon_k  = \varepsilon_F + \hbar v_F k$ linearized 
around the Fermi energy $\varepsilon_F$. 
$b_{k\sigma}$ is the annihilation operator for a conduction electron
with spin $\sigma$ and wavevector $k$ (measured from $k_F$).
The Kondo interaction, controlled by
$J_K$, is an antiferromagnetic coupling between the impurity
spin operator $\vec{S}=\{S^A\}$ and the spin
operator of the conduction electrons at $x=0$ (impurity site).
$T^A$ and $S^A$ are two sets of $N^2-1$ generators satisfying
the commutation relations
\begin{equation}\label{comm}
\vspace*{1.cm}
[ S^A, S^B ] = i f_{ABC} S^C, \quad [ T^A, T^B ] = i f_{ABC} T^C,
\end{equation}
where the antisymmetric tensors $f_{ABC}$ are the structure factors 
of the SU(N) Lie algebra.
The matrices $T^A$ generate the fundamental representation of $SU(N)$ 
while the $S^A$ define the antisymmetric
representation of $SU(N)$ corresponding to a Young tableau of
a single column with $m$ boxes. Physically, the Kondo Hamiltonian~\eqref{kondo}
 emerges from an Anderson
model with exactly $m$ electrons at the impurity site.

In the ground state of the model, the spin of the impurity forms a singlet 
with conduction electrons.
It is therefore completely screened and disappears from the picture
at low energy. 
The Fermi liquid theory describes the low energy regime
and is built on the following assumptions: (i) the singlet
scatters elastically conduction electrons, (ii) virtual polarization
of the singlet leads to weak interactions between  conduction 
electrons of different spin and, (iii) the energy of the system
is an analytical function only of the bare energies $\varepsilon_k$ and of
the relative quasiparticle occupation numbers $\delta n_\sigma (\varepsilon)$.
More precisely, $\delta n_\sigma (\varepsilon) = n_\sigma (\varepsilon)
- \theta (\varepsilon_F - \varepsilon)$ is the actual occupation
number relative to the ground state distribution with
Fermi energy $\varepsilon_F$.
The last point (iii) is in fact the most stringent one and it
is reminiscent of the usual (bulk) Fermi liquid theory.
Instead of considering the total energy, one can concentrate
on the energy shift of a single quasiparticle excitation and, by imposing boundary
condition for a system of finite size,  translate it into 
an electron phase shift  at energy $\varepsilon$.
$\delta_\sigma (\varepsilon,\delta n_{\sigma'})$ is therefore an analytical function
that depends only on $\varepsilon$ and on the functions $\delta n_{\sigma'}
(\varepsilon)$.

The general expansion of the phase shift (hereafter $\sum_\varepsilon$
stands for $\int d \varepsilon$)
\widetext
\begin{equation}\label{pshift}
\begin{split}
\delta_{\sigma} (\varepsilon,\delta n_{\sigma'})
= \delta_0   &+ \frac{\alpha_1}{T_K} (\varepsilon-\varepsilon_F) 
 + \frac{\alpha_2}{T_K^2} (\varepsilon-\varepsilon_F)^2 - \sum_{\sigma'\ne\sigma}
\Bigg( \frac{\phi_1}{T_K} \sum_{\varepsilon'} \delta n_{\sigma'} (\varepsilon') 
\\[1mm] & + \frac{\phi_2}{2 T_K^2}   \sum_{\varepsilon'} ( \varepsilon + \varepsilon' - 2 \varepsilon_F)
\, \delta n_{\sigma'} (\varepsilon') 
- \frac{\chi_2}{ T_K^2} \sum_{\sigma''<\sigma' \atop \sigma'' \ne \sigma} 
\sum_{\varepsilon',\varepsilon^{\prime \prime}}\delta n_{\sigma'} (\varepsilon')
\delta n_{\sigma''} (\varepsilon^{\prime \prime}) \Bigg) 
+ {\cal O} \left( \frac{1}{T_K^3} \right),
\end{split}
\end{equation}
\endwidetext
\noindent introduces the dimensionless phenomenological coefficients $\alpha_1$,
$\alpha_2$, $\phi_1$, $\phi_2$ and $\chi_2$. $\delta_0$ is the phase shift
at the Fermi level. Its value is imposed by the Friedel sum rule,
\begin{equation}\label{zerops}
\delta_0 = \frac{m \, \pi}{N},
\end{equation}
so that $\delta_0 = \pi/2$ at half-filling, {\it i.e.} for a particle-hole
symmetric situation. Only $\alpha_1$ and $\phi_1$
are kept in the conventional FL approach~\cite{nozieres1974a,glazman2005}.  
$\alpha_{1,2}$ correspond to 
elastic scattering. 
$\phi_2$ is an energy correction to the four-point vertex controlled by
$\phi_1$. $\chi_2$ tunes the six-point vertex corresponding to the
local interaction of three electrons.
The properties of the Kondo resonance can be read from the phase shift
expression~\eqref{pshift}. The phase shift expansion
for a resonant level model (RLM) of width~$\sim T_K$ is similar to the first
three (elastic) terms, which identifies $T_K$ as the size of the
Kondo resonance. The comparison with RLM also indicates  that  $\alpha_2$ 
is expected to vanish
when the resonance is centered at the Fermi level~\cite{rajan1983}.
The dependence of the phase shift~\eqref{pshift} on the conduction electron populations
is also physically sensible. The Kondo screening  is a many-body 
effect that results from the sharpness of the Fermi surface~\cite{kondo1964}.
The resonance is therefore extremely sensitive to changes in the 
occupation numbers which modify the shape of the Fermi surface.

\begin{figure}
\includegraphics[width=6.5cm]{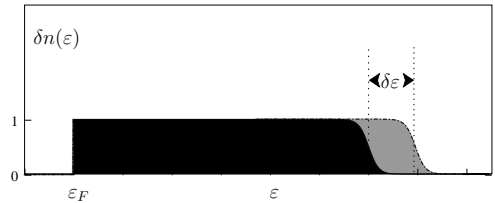}
\caption{\label{shifted} Schematic view of the doping of conduction electrons.
The black filled area represents the initial distribution $\delta n (\varepsilon)$
and the grey one the added electrons $\delta n^1_\sigma (\varepsilon)$.
Both the initial and final ($\delta n'_{\sigma} (\varepsilon)$) 
distributions start at
$\varepsilon = \varepsilon_F$ since the ground state distribution has been
 subtracted.
}
\end{figure}

The {\it floating} of the Kondo resonance follows from the same physical
idea. Since the Kondo resonance is built by the conduction 
electrons themselves, its structure should be invariant when doping
the system such that the shapes of electronic distributions remain the same,
apart from a global energy shift $\delta \varepsilon$. 
The only effect of this doping is then to shift the Kondo resonance by 
 $\delta \varepsilon$. Let us implement this physical idea in a practical way.
The doping procedure is shown in Fig.~\ref{shifted}.
$\delta n'_{\sigma} (\varepsilon)$
denotes the new distribution and $\delta n^1_\sigma (\varepsilon)$ the added
one such that 
\[
\delta n'_{\sigma} (\varepsilon)  = \delta n_{\sigma} (\varepsilon) 
+ \delta n^1_\sigma (\varepsilon).
\]
This translates into $\delta n'_{\sigma} (\varepsilon) =
\delta n_{\sigma} (\varepsilon- \delta \varepsilon) + \theta ( \varepsilon)
- \theta (\varepsilon - \delta \varepsilon)$ since $\delta n'_{\sigma}$
and $\delta n_{\sigma}$ have the same shape at the right of the energy
distribution. The invariance of the Kondo resonance under this doping
implies that
\begin{equation}\label{invariance}
\delta_{\sigma} (\varepsilon + \delta \varepsilon,\delta n'_\sigma)
= \delta_{\sigma} (\varepsilon,\delta n_\sigma),
\end{equation}
for any $\varepsilon$ and $\delta n_\sigma$. Using Eq.~\eqref{pshift},
it leads to four equations 
\begin{subequations}\label{condinva}
\begin{align}
\alpha_1 - (N-1) \phi_1 & = 0, \\[1mm]
\alpha_2 - \frac{3 (N-1)}{4} \phi_2 + \frac{(N-1)(N-2)}{2} \chi_2 &=0, \\[1mm]
(N-2) \chi_2 - \phi_2 = 0,  \qquad 2 \alpha_2 -  \frac{N-1}{2} \phi_2  &=0,
\end{align}
\end{subequations}
corresponding to vanishing coefficients in front of, respectively
$\delta \varepsilon$, $\varepsilon \delta \varepsilon$, 
$\delta \varepsilon \, \sum_{\sigma'\ne \sigma,\varepsilon'} \delta n_{\sigma'} ( \varepsilon')$
and $(\delta \varepsilon)^2$.
Eqs.~\eqref{condinva} are satisfied with
\begin{subequations}\label{FLs}
\begin{align}
\label{FLs1}
\alpha_1 &= (N-1) \phi_1, \\[1mm]\label{FLs2}
\alpha_2 = \frac{N-1}{4} \, \phi_2, & \qquad \phi_2 = (N-2) \, \chi_2.
\end{align}
\end{subequations}
The ratio between $\alpha_1$ and $\phi_1$ was first obtained in Ref.~\cite{nozieres1980}.
The identities~\eqref{FLs} are consistent with the Friedel sum rule but
they cannot be simply reduced to it. It is the whole Kondo resonance structure
that remains invariant through the energy shift and not only the phase shift 
at the Fermi energy.  To our knowledge, this generalization of the original
Nozi\`eres' argument had not yet been pointed out.
Note that the Fermi energy $\varepsilon_F$ is the only
energy reference in this problem, compared to which the system is doped.
An alternative and straightforward way to derive Eqs.~\eqref{condinva} et~\eqref{FLs}
is therefore to require the invariance of the phase shift~\eqref{pshift} 
when shifting $\varepsilon_F$.

\section{Conformal field theory}\label{sec:CFT}

The CFT offers an alternative and illuminating perspective to reexamine these
new FL corrections. It was originally  noted by Affleck~\cite{affleck1990}
 that the fixed ratio 
between elastic and inelastic terms in the leading FL corrections
was a consequence of spin-charge decoupling (spin-charge separation was first
shown in Ref.~\cite{schotte1970}, it also appears in the Bethe ansatz
solutions for the Kondo~\cite{andrei1983+tsvelik1982+rasul1982}
and the Anderson~\cite{schlottmann1982+tsvelik1983} models).
Written in terms of (spin) currents, the only eligible dimension-$3$ operator
is the square of the spin current operator. 
This single operator was shown~\cite{affleck1993},
using standard point-splitting techniques, 
to produce the two couplings in Nozi\`eres' FL theory, thereby enforcing automatically
the relation~\eqref{FLs1}.
We shall see here that the same reduction  applies to the second generation
of FL terms. One single dimension-$4$ operator can be identified, which produces
the couplings $\alpha_2$, $\phi_2$ and $\chi_2$ together with the relations~\eqref{FLs2}.

The quadratic Hamiltonian describing the strong coupling fixed point 
\begin{equation} 
H_0  = \sum_{k,\sigma=1 \ldots N} \varepsilon_k  \psi_{k\sigma}^\dagger \psi_{k\sigma},
\end{equation}
is written in terms of the quasiparticle field $\psi_{\sigma} (x) = \sum_k \psi_{k\sigma} e^{i k x}$.
It corresponds to free fermions and the zero-energy phase shift~\eqref{zerops} is included
in the wavefunction associated to $\psi_{k\sigma}$. The zero-temperature Green's function
is given by 
\begin{equation}\label{green}
\langle \psi^\dagger (x) \psi (x') \rangle = \frac{i}{2 \pi} \frac{1}{x-x'}.
\end{equation}
The spin current operator $J^A (x) = \sum_{\sigma,\sigma'}
\psi^\dagger_\sigma  (x) T^A_{\sigma,\sigma'}
\psi_{\sigma'} (x)$  is written on the basis of SU(N) generators $T^A$.
The $N\times N$ Hermitian and traceless matrices $T^A$ follow
Gell-Mann convention~\cite{azcarraga1998}.
The symmetry tensors $d_{ABC}$~\cite{bensimon2006}
 are defined by the multiplication rules
\begin{equation}
T^A T^B = \frac{1}{2 N} \delta_{A,B} \mathds{1} + \frac{1}{2} (d_{ABC} + i f_{ABC} ) T^C,
\end{equation}
compatible with Eq.~\eqref{comm} and
where $\mathds{1}$ denotes the unit matrix. 
The dimension-$3$ FL correction is given by $H_{\rm I}^{(1)}= - \lambda_1 J^A (0)J^A (0)$.
For the dimension-$4$ operator, 
we seek a SU(N) invariant form involving three spin currents. 
The most natural one is
\begin{equation}\label{pertu}
H_{\rm I}^{(2)} = - \lambda_2  \int d x \, \delta (x) \, d_{ABC} : J^A (x) J^B (x) J^C (x) :,
\end{equation}
which can be seen as a generalization of the cubic Casimir 
operator of the SU(N) Lie algebra~\cite{klein1963}.
The notation $:\ldots:$ indicates normal ordering of the operators.
The invariance over SU(N) rotations can be shown directly using the identity
\[
d_{EBC} \, f_{EDA} + d_{AEC} \, f_{EDB} + d_{ABE} \, f_{EDC} = 0.
\]
The calculation that follows is similar to the one that has been performed for
the dimension-$3$ operator in Refs.~\cite{affleck1993,lehur2007}.
The product $d_{ABC}  J^A J^B J^C $ is obtained from the contraction
of the tensor $d_{ABC}  T_{ab}^A T^B_{cd} T^C_{ef} $ with six fermionic fields 
(here $a,b,c,d,e,f$ denote spins).
We resort to the identity
\begin{equation}\label{ide1}
\begin{split}
& d_{ABC}   T_{ab}^A T^B_{cd} T^C_{ef} = {\cal N} \bigg( \frac{N}{2} ( \delta_{ad} \delta_{be} \delta_{cf}
+ \delta_{af} \delta_{bc} \delta_{de} ) \\[1mm]
& + \frac{2}{N} \delta_{ab} \delta_{cd} \delta_{ef}
- ( \delta_{ab} \delta_{cf} \delta_{de} + \delta_{ad} \delta_{bc} \delta_{ef}
+ \delta_{af} \delta_{be} \delta_{cd})  \bigg),
\end{split}
\end{equation}
with the normalization factor ${\cal N} = (N^2-1)/(2 N (N^2+1))$,
in order to avoid the explicit values of the generators $T^A$. 
The singular operator  $J^A J^B J^C$ is defined using the standard point-splitting
procedure and the normal ordering eventually ensures a regular result.

Using the identity~\eqref{ide1} and the explicit point-splitting calculation -with
the short distance behavior~\eqref{green}-, 
we rewrite the perturbation $H_{\rm I}^{(2)}$~\eqref{pertu} in terms of fermion fields.
This is a tedious but straightforward procedure. The result is proportional to
the combination
\begin{equation}
\begin{split}
 & H_{\rm I}^{(2)} \propto - \frac{2}{3} : \psi^\dagger_{\sigma}  \psi_{\sigma}  
\psi^\dagger_{\sigma'}  \psi_{\sigma'} \psi^\dagger_{\sigma''} \psi_{\sigma''} : \\[1mm]
& + (N-2) \frac{i}{2 \pi} (\partial_1 - \partial_2) : \psi^\dagger_{\sigma,1} \psi_{\sigma,2} 
\psi^\dagger_{\sigma'}  \psi_{\sigma'}  : \\[1mm]
& 
- \frac{(N-2)(N-1)}{4} \left( \frac{i}{2 \pi} \right)^2
 (\partial_1 - \partial_2)^2 : \psi^\dagger_{\sigma,1} \psi_{\sigma,2} : ,
\end{split}
\end{equation}
where all fields are taken at $x=0$. For the complete result, we prefer
to go to wavevector space. Using that $k = 2 \pi \nu_1 \varepsilon_k$,
where $\nu_1 = 1/(h v_F)$ is the density of state for chiral 1D fermions,
it reads
\begin{equation}\label{interaction}
\begin{split}
& H_{\rm I}^{(2)} = - \frac{\alpha_2}{4 \pi \nu_1 T_K^2} \sum_{\sigma,\{ k_i \}} 
(\varepsilon_{k_1} + \varepsilon_{k_2})^2 : \psi_{\sigma,k_1}^\dagger \psi_{\sigma,k_2}
 : \\[2mm]
& + \frac{\phi_2}{\pi \nu_1^2 T_K^2} \sum_{\sigma < \sigma',\{ k_i \}}
 \frac{\sum_{i=1}^4 \varepsilon_{k_i}}{4} 
: \psi_{\sigma,k_1}^\dagger \psi_{\sigma,k_2}
\psi_{\sigma',k_3}^\dagger \psi_{\sigma',k_4}  :  \\[2mm]
& - \frac{\chi_2}{\pi \nu_1^3 T_K^2} \sum_{\sigma < \sigma'<\sigma''\atop\{ k_i \}} 
: \psi_{\sigma,k_1}^\dagger \psi_{\sigma,k_2}
\psi_{\sigma',k_3}^\dagger \psi_{\sigma',k_4} \psi_{\sigma'',k_5}^\dagger \psi_{\sigma'',k_6}  :.
\end{split}
\end{equation}
Together with the dimension-$3$ operators, Eq.~\eqref{interaction} reproduces exactly 
the phase shift~\eqref{pshift}. The coefficients $\alpha_2$, $\phi_2$ and $\chi_2$ 
are related to $\lambda_2$,
\begin{subequations}
\begin{align}
\frac{\alpha_2}{\pi \nu_1 T_K^2} & = 3 \nu_1^2 {\cal N} \frac{(N^2-4)(N^2-1)}{2 N} \, \lambda_2, \\[1mm]
\frac{\phi_2}{\pi \nu_1^2 T_K^2} & = 6  \nu_1 {\cal N} \frac{(N^2-4)(N+1)}{N}\, \lambda_2,\\[1mm]
\frac{\chi_2}{\pi \nu_1^3 T_K^2} & = 6   {\cal N} \frac{(N+2)(N+1)}{N}\, \lambda_2,
\end{align}
\end{subequations}
so that again we find the Eqs.~\eqref{FLs2}.

To conclude, we have found independently that the CFT leads to the 
same three corrections with the same
relations~\eqref{FLs2} as the FL theory.

\section{Input from the Bethe ansatz}\label{sec:ba}

In the last two sections, we have shown that we can relate the amplitudes of 
the different physical processes that appear at a given order in the
Hamiltonian perturbative expansion. This can be done either in the FL or
in the CFT framework. The arguments that we have used are only based 
on symmetries and on the global structure of the low energy resonance.
What we cannot do however with these phenomenological approaches is to relate the 
 coefficients of the different orders, for instance $\alpha_1$ to $\alpha_2$ or
similarly $\lambda_1$ to $\lambda_2$. For this, we have to resort to the
exact solution of the model, in principle given by the Bethe ansatz
solution. Using an alternative Bethe ansatz technique, Bazhanov, Lukyanov
and Tsvelik~\cite{bazhanov2003} have derived analytical expressions for the free energy 
in the general SU(N) Kondo model with $m$ electrons forming the impurity.
We can compute the free energy perturbatively with our model and then
compare with the exact solution as a way 
to extract the relationship between $\alpha_1$ and $\alpha_2$.

We study the same situation as in Ref.~\cite{bazhanov2003}. The system is at zero temperature and
independent generalized magnetic fields $h_\sigma$ are applied to the different
spin components. Their chemical potentials  are then shifted
to $\epsilon_F + h_\sigma$ (or $h_\sigma$ alone if we take $\epsilon_F=0$).
Since the position of $\epsilon_F$ is arbitrary as we have demonstrated in Sec.~\ref{sec:FL},
it is chosen such that $\sum_\sigma h_\sigma = 0$.
In the FL theory, the free energy is straightforward to calculate from the phase shift~\eqref{pshift},
\begin{equation}\label{enerFL}
\begin{split}
F & = F_0  - \frac{1}{\pi T_K} \sum_{\sigma,\varepsilon} \left( \alpha_1 \varepsilon + \frac{\alpha_2}{T_K}  \varepsilon^2 \right)
\delta n_\sigma (\varepsilon) \\[1mm]
& + \frac{1}{\pi T_K} \sum_{\sigma < \sigma' \atop \varepsilon, \varepsilon' } \left( \phi_1 + \frac{\phi_2}{T_K}
\frac{\varepsilon + \varepsilon'}{2} \right)\delta n_\sigma (\varepsilon)
\delta n_{\sigma'} (\varepsilon') \\[1mm]
& - \frac{\chi_2}{\pi T_K^2} \sum_{\sigma < \sigma' <\sigma'' \atop \varepsilon, \varepsilon',\varepsilon''}
\delta n_\sigma (\varepsilon)
\delta n_{\sigma'} (\varepsilon') 
\delta n_{\sigma''} (\varepsilon''),
\end{split}
\end{equation}
where $F_0$ is the ground state energy.
The same expression can be recovered from the Hamiltonian form, with~\cite{glazman2005,sela2006+golub2006+gogolin2006b}
\begin{equation}\label{firstcorr}
\begin{split}
& H_{\rm I}^{(1)} = - \frac{\alpha_1}{2 \pi \nu_1 T_K} \sum_{\sigma,\{ k_i \}} 
(\varepsilon_{k_1} + \varepsilon_{k_2}) : \psi_{\sigma,k_1}^\dagger \psi_{\sigma,k_2}
 : \\[2mm]
& + \frac{\phi_1}{\pi \nu_1^2 T_K} \sum_{\sigma < \sigma',\{ k_i \}}
: \psi_{\sigma,k_1}^\dagger \psi_{\sigma,k_2}
\psi_{\sigma',k_3}^\dagger \psi_{\sigma',k_4}  : ,
\end{split}
\end{equation}
and Eq.~\eqref{interaction}.
The free energy expression~\eqref{enerFL} is general. In our simple case, the energy integrals are
easy to perform with $\delta n_\sigma (\varepsilon) = \theta(\varepsilon) - \theta(\varepsilon - h_\sigma)$.
Using the FL relations~\eqref{FLs}, the final result is
\begin{equation}
F = F_0 - A_1 \left( \sum_\sigma h_\sigma^2 \right)
- A_2 \left( \sum_\sigma h_\sigma^3 \right),
\end{equation}
with the coefficients $A_1 = \frac{N \alpha_1}{2 \pi T_K (N-1)}$ and
$A_2 = \frac{\alpha_2}{3 \pi T_K^2} \frac{N^2}{(N-1)(N-2)}$. On the other hand, the
exact formula~\cite{bazhanov2003} gives
$A_1 = \sin (m \pi /N)/ ( 2 \pi T_K \sin (\pi/N) )$ and 
\[
A_2 = \frac{\sin(2 m \pi /N)}{\sin (2 \pi/N)} \frac{\Gamma(1/N)}{\Gamma(1/2+1/N) 3 \pi^{3/2} T_K^2},
\]
with the gamma function $\Gamma (z)$.
The following universal ratio can be extracted
\begin{equation}\label{FLs3}
\frac{\alpha_2}{\alpha_1^2} = \frac{N-2}{N-1} \, \frac{\Gamma(1/N)}{\sqrt{\pi}\Gamma\left(\frac{1}{2} + \frac{1}{N} \right)}
\frac{\tan(\pi/N)}{\tan( m \pi/N)}.
\end{equation}
With this relation and the Eqs.~\eqref{FLs}, all coefficients of the model are related to $\alpha_1$
and our low energy approach is fully characterized. Note that the precise value of $\alpha_1$ 
depends on the definition of the Kondo temperature. With no loss of generality, 
we can set $\alpha_1=1$
and $T_K$ is the only energy scale that controls the low energy expansion.

For a half-filled dot (particle-hole symmetric case) like the standard SU(2) case, $m = N/2$ so that
$\alpha_2 = 0$ from Eq.~\eqref{FLs3}, and $\phi_2 = \chi_2 = 0$ from Eqs.~\eqref{FLs2}.
This indicates, as we have already mentioned, that the Kondo resonance is centered exactly 
at the Fermi level as a natural consequence of particle-hole symmetry.
Another interesting case is the large $N$ limit of Eq.~\eqref{FLs3}.
In this limit, the Kondo model becomes a resonant level model with
a position and a width that are determined in a mean-field way (the slave boson mean field
theory~\cite{delattre2009,coleman1984,newns1987}). 
For $N \to +\infty$, we indeed find that Eq.~\eqref{FLs3} tends to 
$\alpha_2/\alpha_1^2 \simeq \cot (\delta_0)$ - with $\delta_0$ given by Eq.~\eqref{zerops} - 
as expected for  a resonant level model.

\section{Comparison with 1/N expansion}\label{sec:1n}

The extended FL theory that we have built allows us to compute
observables in the low energy regime.
A Hamiltonian form is used for the perturbing operators,
given by Eqs.~\eqref{interaction} and~\eqref{firstcorr},
and electron interaction is incorporated by standard
many-body diagrammatics. 
Following the large $N$ approach developed by  Read and Newns~\cite{newns1987},
Houghton, Read and Won~\cite{houghton1987} have calculated 
 the conductivity and the Lorentz ratio at low energy and to first order 
in a systematic $1/N$ expansion. We shall next compute these
 transport properties in the same limit and see that our analytical predictions
coincide exactly  with those of Ref.~\cite{houghton1987}.

We consider the conventional Kondo problem~\cite{affleck1993,houghton1987}: 
a host metal with density
of state  $\nu_3$ at the Fermi energy contains dilute SU(N) Kondo impurities
with density $n_i$. The single-particle lifetime $\tau (\varepsilon,T)$ for conduction
electrons is related
to the imaginary part of the 1D improper self-energy (see Ref.~\cite{affleck1993} for
more details),
\begin{equation}
\frac{1}{\tau (\varepsilon,T)} = - \frac{2 n_i}{\nu_3} \, \, {\rm Im} \Sigma_1^R (\varepsilon,T).
\end{equation}
The different moments of $\tau$ can be defined as
\begin{equation}
{\cal L}^n (T) = \int_{-\infty}^{+\infty} d \varepsilon 
\left(- \frac{\partial f(\varepsilon,T)}{\partial \varepsilon} \right) 
\tau (\varepsilon,T) \varepsilon^n,
\end{equation}
where $f(\varepsilon,T) = (1+ e^{\varepsilon/T})^{-1}$ is the  finite temperature
Fermi-Dirac distribution.
The conductivity and the Lorentz ratio are then respectively given by~\cite{houghton1987}
\begin{subequations}\label{obs}
\begin{align}
\sigma (T) &= \frac{\nu_3 e^2 v_F^2}{3} \, {\cal L}^0 (T), \\[1mm]
\frac{L(T)}{L_0} &= \frac{3}{(\pi T)^2} \, \left[ \frac{{\cal L}^2}{{\cal L}^0}
- \left( \frac{{\cal L}^1}{{\cal L}^0} \right)^2 \right],
\end{align}
\end{subequations}
with $L_0 = \pi^2/3 e^2$. The Lorentz ratio is defined as $L = \kappa/ \sigma T$
where $\kappa$ is the thermal conductivity.

\begin{figure}
\includegraphics[width=5.5cm]{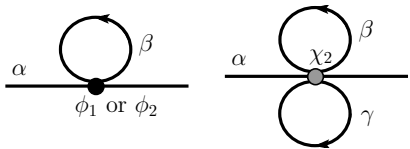}
\caption{\label{har-diag} Hartree diagrams for the self-energy built 
from Eqs.~\eqref{interaction} and~\eqref{firstcorr}. The full dots (resp. black and grey)
indicate vertices with four or six external
lines. $\alpha$, $\beta$ and $\gamma$ denote spins.}
\end{figure}

\begin{figure}
\includegraphics[width=4.5cm]{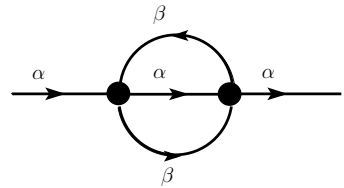}
\caption{\label{collision}  Second order contribution to the self-energy
corresponding to inelastic collisions.}
\end{figure}

We gather all terms that contribute to  the self-energy $\Sigma_1^R$ up to ${\cal O}(1/T_K^2)$.
Following Ref.~\cite{affleck1993}, the elastic contributions 
can be summed up to give
\begin{equation}\label{elastic}
\Sigma^R_{1,{\rm el}} (\varepsilon,T) = - \frac{i}{2 \pi} \left( 1 - e^{2 i \delta_{\rm el}( \varepsilon )}
\right),
\end{equation}
where $\delta_{\rm el} ( \varepsilon ) = \delta_0 + (\alpha_1/T_K) \varepsilon +
(\alpha_2/T_K^2) \varepsilon^2$ is the elastic phase shift. As in Ref.~\cite{houghton1987}, 
the impurity is formed by only one electron so that $\delta_0 = \pi/N$.
We next turn to electron interaction. The Hartree diagrams, shown in Fig.~\ref{har-diag},
have a structure similar to potential scattering. 
Therefore they can be incorporated
into the elastic expression~\eqref{elastic} where the phase shift is now
given by Eq.~\eqref{pshift} with $\delta n (\varepsilon) = f (\varepsilon,T) 
-\theta (\varepsilon_F- \varepsilon)$.
 More precisely, since 
$\sum_{\varepsilon} \delta n (\varepsilon) = 0$ and $\sum_{\varepsilon} 
\varepsilon \, \delta n (\varepsilon) = (\pi T)^2/6$, the phase shift~\eqref{pshift} simplifies to
\begin{equation}\label{pshift2}
\delta(\varepsilon,T) = \delta_{\rm el} ( \varepsilon ) - \frac{(N-1) \phi_2}{T_K^2} \frac{(\pi T)^2}{12}, 
\end{equation}
where only the $\phi_2$ coupling survives.
The last diagram to consider is shown Fig.~\ref{collision}. 
It describes
relaxation due to electron inelastic collisions~\cite{nozieres1974a,nozieres1974b+nozieres1978}. 
Its calculation follows from Ref.~\cite{affleck1993} leading
to
\begin{equation}\label{inelastic}
\Sigma^R_{1,{\rm in}} (\varepsilon,T) = - \frac{i \, e^{2 i \delta_0}}{4} (N-1) \left(
\frac{\phi_1}{T_K} \right)^2 \left[ \varepsilon^2 + (\pi T)^2 \right],
\end{equation}
where the $N-1$ factor comes from the intermediate spin summation.

To summarize our findings, the self-energy $\Sigma_1^R = \Sigma^R_{1,{\rm el}}
+ \Sigma^R_{1,{\rm in}} $ is the sum of inelastic~\eqref{inelastic} and
 elastic~\eqref{elastic} contributions with the phase shift~\eqref{pshift2}.
From this result, the transport observables~\eqref{obs} can be determined
at low energy for any $N$. Instead, we start at this point 
to investigate the large $N$ limit keeping only the first order $1/N$ corrections.
Hence we approximate $\sin \delta_0 \simeq \pi/ N$ and $\cos 2 \delta_0 \simeq 1$.
Expanding the single-particle lifetime at low energy, we find
\begin{equation}\label{tau}
\begin{split}
& \frac{\tau (\varepsilon,T)}{\tau(0,0)} = 1 - \frac{2 \bar{\alpha}_1 \varepsilon}{T_K}
 + \frac{3 \bar{\alpha}_1^2 \varepsilon^2}{T_K^2}  \\[1mm]
& - \frac{2 \bar{\alpha}_2 }{T_K^2}
\left[\varepsilon^2 - \frac{(\pi T)^2}{3} \right]
- \frac{\bar{\alpha}_1^2}{T_K^2} \frac{1}{2 N} [\varepsilon^2 + (\pi T)^2 ],
\end{split}
\end{equation}
with the renormalized coefficients $\bar{\alpha}_{1,2} = (N / \pi) \alpha_{1,2}$.
Before proceeding further, 
let us discuss the normalization of $\alpha_1$. The Kondo temperatures in the
FL theory and in the large $N$ approach of Ref.~\cite{houghton1987} coincide 
if  a single observable is matched between the two models,
for instance  the zero temperature magnetic susceptibility.
In the FL theory, it reads~\cite{nozieres1974a,nozieres1974b+nozieres1978,nozieres1980}
\begin{equation}
\chi_{0} = \frac{N \alpha_1}{N-1}  \frac{N(N^2-1)}{12} \frac{(g \mu_B)^2}{\pi T_K},
\end{equation}
whereas $\frac{1}{3} (g \mu_B)^2 J(J+1)/T_K$ is the definition given 
in Ref.~\cite{houghton1987}.
 $J$ is the angular momentum and the impurity model has $SU(2J+1)$ symmetry.
A common Kondo temperature $T_K$ is thus achieved with $\bar{\alpha_1} = 1-1/N$.

The conductance~\eqref{obs} is readily obtained from the electron lifetime~\eqref{tau}
with the result
\begin{equation}\label{cond}
\begin{split}
\frac{\sigma(T)}{\sigma(0)} & =  1+ \left( \frac{\pi T}{T_K} \right)^2
\left[ \bar{\alpha}_1^2 \left( 1 - \frac{2}{3 N} \right) + {\cal O} 
\left(\frac{1}{N^2} \right)   \right] \\[1mm]
& =  1+ \left( \frac{\pi T}{T_K} \right)^2 
\left[  1 - \frac{8}{3 N}  + {\cal O} 
\left(\frac{1}{N^2} \right)   \right]   ,
\end{split}
\end{equation}
in full agreement with Ref.~\cite{houghton1987}. This agreement 
confirms that the two procedures, namely the FL theory expanded at large $N$ 
on one side, and the large $N$ approach expanded at low 
energy on the other side, indeed correspond to the same physical limit.
Nevertheless it does not help us to validate 
the new dimension-$4$ FL corrections since $\bar{\alpha}_2$ 
disappears from the final result~\eqref{cond}.

The situation is markedly different for the Lorentz ratio~\eqref{obs}. 
Using
\[ 
\int_{-\infty}^{-\infty} d \varepsilon \, \varepsilon^4  
\left(- \frac{\partial f(\varepsilon,T)}{\partial \varepsilon} \right)
= \frac{7 (\pi T)^4}{15},
\]
and the electron lifetime~\eqref{tau}, we obtain, up to ${\cal O} (1/N)$, 
\begin{equation}\label{loren}
\frac{L(T)}{L_0} =  1+ \frac{8}{15} \left( \frac{\pi T}{T_K} \right)^2
\left[ \frac{7}{2} \bar{\alpha}_1^2 - 4  \bar{\alpha}_2^2 - \frac{1}{N}   \right],
\end{equation}
where $\bar{\alpha}_2$ is explicitly present. The large $N$ expansion
of the universal ratio~\eqref{FLs3} (with $m=1$),
\begin{equation}
\frac{\alpha_2}{\alpha_1^2} \simeq \frac{N}{\pi} \, \left( 1 + \frac{2 \ln 2 -1}{N} \right),
\end{equation}
is introduced in Eq.\eqref{loren}, leading eventually to
\begin{equation}\label{loren2}
\frac{L(T)}{L_0} =  1 - \frac{4}{15} \left( \frac{\pi T}{T_K} \right)^2
\left[ 1 +  \frac{8}{N} ( 2 \ln 2 -1)   \right].
\end{equation}
Again there is full agreement with Ref.\cite{houghton1987}.

One conclusion that can be drawn from these results is that our extension
of the FL theory satisfies a stringent test imposed by the large $N$ approach.
We can also be confident in our theory and reverse the perspective with
the following conclusion: we have checked on representative observables 
that the $1/N$ expansion of Read and Newns is correct at low energy.

\section{Conclusions}\label{sec:c}

In the case of a generalized SU(N) symmetry for the impurity away from half-filling, 
the Kondo resonance is centered off the Fermi energy. One consequence is that observables
like the resistivity in magnetic alloys, or the current and the noise in quantum dots,
require at low energy the introduction of the next-to-leading order correction
around the Fermi liquid fixed point.
Two possible reasonings have been employed in this work to identify the new
Fermi liquid corrections. In a first approach, the Landau expansion of the phase
shift has been pushed to the next order. The coefficients of the three
resulting new contributions have further been related by using the {\it floating}
argument. Physically, the {\it floating} expresses the fact that the Kondo 
resonance is built only by the distribution of conduction electron and follows its
Fermi singularity.

In a second approach, we have proposed a single operator, cubic in the spin currents,
and which remains invariant over SU(N) rotations.
This operator resembles the cubic Casimir invariant of the SU(N) Lie algebra.
Performing point  splitting, we have recovered the same three processes with the same
relation between their coupling constants.  In fact, the reduction of coupling constants
can be assigned to a common physical origin: the quenching of charge excitation on the
impurity.
In the first approach, the only fixed {\it absolute} energy reference that the Kondo
resonance might depend on, is the single-particle energy level. It is effectively
pushed to infinity in the Kondo limit which allows to develop the {\it floating}
argument. In the second approach, the fact that charge 
excitations are frozen imposes that our cubic operator involves only spin currents.

Next the ratio between the leading and the next-to-leading order corrections
has been determined by comparison with the exact solution for the free energy.
This reduces further the number of coupling constants to a single one which
is essentially the inverse of the Kondo temperature. Finally, the large $N$
regime of our theory has been shown to coincide exactly with field theoretical
large $N$ predictions, thereby comforting our analysis.

Let us conclude by noting some consequences for experiments (experiments
in alloys with magnetic impurities 
are reviewed in Ref.~\cite{schlottmann1989} with a comparison to exact
Bethe ansatz results).
The subtleties of this work do not apply to the ordinary spin-$1/2$ Kondo
effect with SU(2) symmetry since our novel corrections all vanish in that
case (and for a half-filled dot in general). However, for experiments
probing a possible SU(4) Kondo effect, the ingredients presented here
are necessary to determine the low energy properties of the model.

The author is grateful to M. Bauer, M.-S. Choi, A. A. Clerk, K. Le Hur, T. Kontos, 
X. Leyronas and N. Regnault for interesting discussions.

\newcommand{{{\PRB}}}{{{Phys. Rev. B}}}\newcommand{{{\PR}}}{{{Phys. Rev.}}}\newcommand{{{\PRA}}}{{{Phys. Rev. A}}}\newcommand{{{\PRL}}}{{{Phys. Rev. Lett}}}\newcommand{{{\NPB}}}{{{Nucl. Phys. B}}}\newcommand{{{\RMP}}}{{{Rev. Mod. Phys.}}}\newcommand{{{\ADV}}}{{{Adv. Phys.}}}\newcommand{{{\EPJB}}}{{{Eur. Phys. J. B}}}\newcommand{{{\EPJD}}}{{{Eur. Phys. J. D}}}\newcommand{{{\JPSJ}}}{{{J. Phys. Soc. Jpn.}}}\newcommand{{{\JLTP}}}{{{J. Low Temp. Phys.}}}\newcommand{{{\PTP}}}{{{Progr. Theoret. Phys.}}}\newcommand{{{\PTPS}}}{{{Progr. Theoret. Phys. Suppl.}}}


\begin{thebibliography}{79}
\expandafter\ifx\csname natexlab\endcsname\relax\def\natexlab#1{#1}\fi
\expandafter\ifx\csname bibnamefont\endcsname\relax
\def\bibnamefont#1{#1}\fi
\expandafter\ifx\csname bibfnamefont\endcsname\relax
\def\bibfnamefont#1{#1}\fi
\expandafter\ifx\csname citenamefont\endcsname\relax
\def\citenamefont#1{#1}\fi
\expandafter\ifx\csname url\endcsname\relax
\def\url#1{\texttt{#1}}\fi
\expandafter\ifx\csname urlprefix\endcsname\relax\def\urlprefix{URL }\fi
\providecommand{\bibinfo}[2]{#2}
\providecommand{\eprint}[2][]{\url{#2}} 
\bibitem{hewson1993}\bibinfo{author}{\bibfnamefont{A. Hewson}}, \emph{\bibinfo{title}{The Kondo Problem to Heavy Fermions} }(\bibinfo{publisher}{Cambridge University Press, Cambridge} \bibinfo{year}{1993}).
\bibitem{oguri2001+meir2002+rosch2003+kehrein2005+mehta2006+anders2008} \bibinfo{author}{\bibfnamefont{A. Oguri}}, \bibinfo{journal}{\PRB} \textbf{\bibinfo{volume}{64} }\bibinfo{page}{153305} (\bibinfo{year}{2001}); \bibinfo{author}{\bibfnamefont{Y. Meir}} and \bibinfo{author}{\bibfnamefont{A. Golub}}, \bibinfo{journal}{\PRL} \textbf{\bibinfo{volume}{88} }\bibinfo{page}{116802} (\bibinfo{year}{2002}); \bibinfo{author}{\bibfnamefont{A. Rosch, J. Paaske, J. Kroha, P. W\"olfle}}, \bibinfo{journal}{\PRL} \textbf{\bibinfo{volume}{90} }\bibinfo{page}{076804} (\bibinfo{year}{2003}); \bibinfo{author}{\bibfnamefont{S. Kehrein}}, \bibinfo{journal}{\PRL} \textbf{\bibinfo{volume}{95} }\bibinfo{page}{056602} (\bibinfo{year}{2005}); \bibinfo{author}{\bibfnamefont{P. Mehta, N. Andrei}}, \bibinfo{journal}{\PRL} \textbf{\bibinfo{volume}{96} }\bibinfo{page}{216802} (\bibinfo{year}{2006}); \bibinfo{author}{\bibfnamefont{F. B. Anders}}, \bibinfo{journal}{\PRL} \textbf{\bibinfo{volume}{101} }\bibinfo{page}{066804} (\bibinfo{year}{2008}).
\bibitem{nozieres1974a}\bibinfo{author}{\bibfnamefont{P. Nozi\`eres}}, \bibinfo{journal}{\JLTP} \textbf{\bibinfo{volume}{17} }\bibinfo{page}{31} (\bibinfo{year}{1974}).
\bibitem{nozieres1974b+nozieres1978} \bibinfo{author}{\bibfnamefont{P. Nozi\`eres}}, in \emph{\bibinfo{booktitle}{Proceedings of the 14th International Conference on Low Temperature Physics} }edited by \bibinfo{editor}{M. Krasius and M. Vuorio} (\bibinfo{publisher}{North Holland, Amsterdam,} \bibinfo{year}{1974}) \bibinfo{note}{pp. 339-374}; \bibinfo{author}{\bibfnamefont{P. Nozi\`eres}}, \bibinfo{journal}{J. Physique} \textbf{\bibinfo{volume}{39} }\bibinfo{page}{1117} (\bibinfo{year}{1978}).
\bibitem{wilson1975}\bibinfo{author}{\bibfnamefont{K.G. Wilson}}, \bibinfo{journal}{\RMP} \textbf{\bibinfo{volume}{47} }\bibinfo{page}{773} (\bibinfo{year}{1975}).
\bibitem{affleck1990}\bibinfo{author}{\bibfnamefont{I. Affleck}}, \bibinfo{journal}{\NPB} \textbf{\bibinfo{volume}{336} }\bibinfo{page}{517} (\bibinfo{year}{1990}).
\bibitem{affleck1993}\bibinfo{author}{\bibfnamefont{I. Affleck}} and \bibinfo{author}{\bibfnamefont{A. W. W. Ludwig}}, \bibinfo{journal}{\PRB} \textbf{\bibinfo{volume}{48} }\bibinfo{page}{7297} (\bibinfo{year}{1993}).
\bibitem{affleck1991}\bibinfo{author}{\bibfnamefont{I. Affleck, A. W. W. Ludwig}}, \bibinfo{journal}{\NPB} \textbf{\bibinfo{volume}{352} }\bibinfo{page}{849} (\bibinfo{year}{1991}).
\bibitem{lesage1999a+lesage1999b} \bibinfo{author}{\bibfnamefont{F. Lesage, H. Saleur}}, \bibinfo{journal}{\PRL} \textbf{\bibinfo{volume}{82} }\bibinfo{page}{4540} (\bibinfo{year}{1999}); \bibinfo{author}{\bibfnamefont{F. Lesage, H. Saleur}}, \bibinfo{journal}{\NPB} \textbf{\bibinfo{volume}{546} }\bibinfo{page}{585} (\bibinfo{year}{1999}).
\bibitem{yosida1970+yamada1975a+yosida1975+yamada1975b} \bibinfo{author}{\bibfnamefont{K. Yosida, K. Yamada}}, \bibinfo{journal}{\PTPS} \textbf{\bibinfo{volume}{46} }\bibinfo{page}{244} (\bibinfo{year}{1970}); \bibinfo{author}{\bibfnamefont{K. Yamada}}, \bibinfo{journal}{\PTP} \textbf{\bibinfo{volume}{53} }\bibinfo{page}{970} (\bibinfo{year}{1975}); \bibinfo{author}{\bibfnamefont{K. Yosida, K. Yamada}}, \bibinfo{journal}{\PTP} \textbf{\bibinfo{volume}{53} }\bibinfo{page}{1286} (\bibinfo{year}{1975}); \bibinfo{author}{\bibfnamefont{K. Yamada}}, \bibinfo{journal}{\PTP} \textbf{\bibinfo{volume}{54} }\bibinfo{page}{316} (\bibinfo{year}{1975}).
\bibitem{hewson1994}\bibinfo{author}{\bibfnamefont{A.C. Hewson}}, \bibinfo{journal}{Adv. Phys.} \textbf{\bibinfo{volume}{43} }\bibinfo{page}{543} (\bibinfo{year}{1994}).
\bibitem{nozieres1980}\bibinfo{author}{\bibfnamefont{P. Nozi\`eres, A. Blandin}}, \bibinfo{journal}{J. Phys. (Paris)} \textbf{\bibinfo{volume}{41} }\bibinfo{page}{193} (\bibinfo{year}{1980}).
\bibitem{coqblin1969}\bibinfo{author}{\bibfnamefont{B. Coqblin, J.R. Schrieffer}}, \bibinfo{journal}{\PR} \textbf{\bibinfo{volume}{185} }\bibinfo{page}{847} (\bibinfo{year}{1969}).
\bibitem{andrei1983+tsvelik1982+rasul1982} \bibinfo{author}{\bibfnamefont{N. Andrei, K. Furuya, J.H. Lowenstein}}, \bibinfo{journal}{\RMP} \textbf{\bibinfo{volume}{55} }\bibinfo{page}{331} (\bibinfo{year}{1983}); \bibinfo{author}{\bibfnamefont{A.M. Tsvelik, P.B. Wiegmann}}, \bibinfo{journal}{J. Phys. C} \textbf{\bibinfo{volume}{15} }\bibinfo{page}{1707} (\bibinfo{year}{1982}); \bibinfo{author}{\bibfnamefont{J.W. Rasul}}, in \emph{\bibinfo{booktitle}{Valence Instabilities} }edited by \bibinfo{editor}{P. Wachter and H. Boppart} (\bibinfo{publisher}{North Holland, Amsterdam,} \bibinfo{year}{1982}) \bibinfo{note}{p. 49}.
\bibitem{schlottmann1982+tsvelik1983} \bibinfo{author}{\bibfnamefont{P. Schlottmann}}, \bibinfo{journal}{Z. Phys. B} \textbf{\bibinfo{volume}{49} }\bibinfo{page}{109} (\bibinfo{year}{1982}); \bibinfo{author}{\bibfnamefont{A.M. Tsvelik, P.B. Wiegmann}}, \bibinfo{journal}{\ADV} \textbf{\bibinfo{volume}{32} }\bibinfo{page}{453} (\bibinfo{year}{1983}).
\bibitem{sasaki2004}\bibinfo{author}{\bibfnamefont{S. Sasaki, S. Amaha, N. Asakawa, M. Eto,}} and \bibinfo{author}{\bibfnamefont{S. Tarucha}}, \bibinfo{journal}{} \textbf{\bibinfo{volume}{93} }\bibinfo{page}{017205} (\bibinfo{year}{2004}).
\bibitem{herrero2005+makarovski2007a+makarovski2007b} \bibinfo{author}{\bibfnamefont{P. Jarillo-Herrero, J. Kong, H. S. van der Zant, C. Dekker, L. P. Kouwenhoven, S. D. Franceschi}}, \bibinfo{journal}{Nature(London)} \textbf{\bibinfo{volume}{434} }\bibinfo{page}{484} (\bibinfo{year}{2005}); \bibinfo{author}{\bibfnamefont{A. Makarovski, A. Zhukov, J. Liu, G. Finkelstein}}, \bibinfo{journal}{\PRB} \textbf{\bibinfo{volume}{75} }\bibinfo{page}{241407} (\bibinfo{year}{2007}); \bibinfo{author}{\bibfnamefont{A. Makarovski, J. Liu, G. Finkelstein}}, \bibinfo{journal}{\PRL} \textbf{\bibinfo{volume}{99} }\bibinfo{page}{066801} (\bibinfo{year}{2007}).
\bibitem{delattre2009}\bibinfo{author}{\bibfnamefont{T. Delattre et al.}}, \bibinfo{journal}{Nature Physics} \textbf{\bibinfo{volume}{5} }\bibinfo{page}{208} (\bibinfo{year}{2009}).
\bibitem{langreth1966}\bibinfo{author}{\bibfnamefont{D.C. Langreth}}, \bibinfo{journal}{\PR} \textbf{\bibinfo{volume}{150} }\bibinfo{page}{516} (\bibinfo{year}{1966}).
\bibitem{bazhanov2003}\bibinfo{author}{\bibfnamefont{V.V. Bazhanov, S.L. Lukyanov, A.M. Tsvelik}}, \bibinfo{journal}{\PRB} \textbf{\bibinfo{volume}{68} }\bibinfo{page}{094427} (\bibinfo{year}{2003}).
\bibitem{glazman2005}\bibinfo{author}{\bibfnamefont{L.I. Glazman}} and \bibinfo{author}{\bibfnamefont{M. Pustilnik}}, in \emph{\bibinfo{booktitle}{Nanophysics: Coherence and Transport} }edited by \bibinfo{editor}{H. Bouchiat et al.} (\bibinfo{publisher}{Elsevier} \bibinfo{year}{2005}) \bibinfo{note}{pp. 427-478, arXiv:cond-mat/0501007}.
\bibitem{vitu2008}\bibinfo{author}{\bibfnamefont{P. Vitushinsky, A. A. Clerk,}} and \bibinfo{author}{\bibfnamefont{K. Le Hur}}, \bibinfo{journal}{\PRL} \textbf{\bibinfo{volume}{100} }\bibinfo{page}{036603} (\bibinfo{year}{2008}).
\bibitem{lehur2007}\bibinfo{author}{\bibfnamefont{K. Le Hur, P. Simon,}} and \bibinfo{author}{\bibfnamefont{D. Loss}}, \bibinfo{journal}{\PRB} \textbf{\bibinfo{volume}{75} }\bibinfo{page}{035332} (\bibinfo{year}{2007}).
\bibitem{mora2008}\bibinfo{author}{\bibfnamefont{C. Mora, X. Leyronas, N. Regnault}}, \bibinfo{journal}{\PRL} \textbf{\bibinfo{volume}{100} }\bibinfo{page}{036604} (\bibinfo{year}{2008}).
\bibitem{mora2009}\bibinfo{author}{\bibfnamefont{C. Mora, X. Leyronas, N. Regnault}}, \bibinfo{journal}{\PRL} \textbf{\bibinfo{volume}{102} }\bibinfo{page}{139902} (\bibinfo{year}{2009}).
\bibitem{newns1987}\bibinfo{author}{\bibfnamefont{D. M. Newns, N. Read}}, \bibinfo{journal}{Adv. Phys.} \textbf{\bibinfo{volume}{36} }\bibinfo{page}{799} (\bibinfo{year}{1987}).
\bibitem{rajan1983}\bibinfo{author}{\bibfnamefont{V. T. Rajan}}, \bibinfo{journal}{\PRL} \textbf{\bibinfo{volume}{51} }\bibinfo{page}{308} (\bibinfo{year}{1983}).
\bibitem{kondo1964}\bibinfo{author}{\bibfnamefont{J. Kondo}}, \bibinfo{journal}{\PTP} \textbf{\bibinfo{volume}{32} }\bibinfo{page}{37} (\bibinfo{year}{1964}).
\bibitem{schotte1970}\bibinfo{author}{\bibfnamefont{K. D. Schotte}}, \bibinfo{journal}{Z. Physik} \textbf{\bibinfo{volume}{230} }\bibinfo{page}{99} (\bibinfo{year}{1970}).
\bibitem{azcarraga1998}\bibinfo{author}{\bibfnamefont{J. A. de Azc\'arraga, A. J. Macfarlane, A. J. Mountain, J. C. Pérez Bueno}}, \bibinfo{journal}{\NPB} \textbf{\bibinfo{volume}{510} }\bibinfo{page}{657} (\bibinfo{year}{1998}).
\bibitem{bensimon2006}\bibinfo{author}{\bibfnamefont{D. Bensimon, A. Jerez, M. Lavagna}}, \bibinfo{journal}{\PRB} \textbf{\bibinfo{volume}{73} }\bibinfo{page}{224445} (\bibinfo{year}{2006}).
\bibitem{klein1963}\bibinfo{author}{\bibfnamefont{A. Klein}}, \bibinfo{journal}{J. Math. Phys.} \textbf{\bibinfo{volume}{4} }\bibinfo{page}{1283} (\bibinfo{year}{1963}).
\bibitem{sela2006+golub2006+gogolin2006b} \bibinfo{author}{\bibfnamefont{E. Sela, Y. Oreg, F. von Oppen,}} and \bibinfo{author}{\bibfnamefont{J. Koch}}, \bibinfo{journal}{\PRL} \textbf{\bibinfo{volume}{97} }\bibinfo{page}{086601} (\bibinfo{year}{2006}); \bibinfo{author}{\bibfnamefont{A. Golub}}, \bibinfo{journal}{\PRB} \textbf{\bibinfo{volume}{73} }\bibinfo{page}{233310} (\bibinfo{year}{2006}); \bibinfo{author}{\bibfnamefont{A.O. Gogolin}} and \bibinfo{author}{\bibfnamefont{A. Komnik}}, \bibinfo{journal}{\PRL} \textbf{\bibinfo{volume}{97} }\bibinfo{page}{016602} (\bibinfo{year}{2006}).
\bibitem{coleman1984}\bibinfo{author}{\bibfnamefont{P. Coleman}}, \bibinfo{journal}{\PRB} \textbf{\bibinfo{volume}{29} }\bibinfo{page}{3035} (\bibinfo{year}{1984}).
\bibitem{houghton1987}\bibinfo{author}{\bibfnamefont{A. Houghton, N. Read, H. Won}}, \bibinfo{journal}{\PRB} \textbf{\bibinfo{volume}{35} }\bibinfo{page}{5123} (\bibinfo{year}{1987}).
\bibitem{schlottmann1989}\bibinfo{author}{\bibfnamefont{P. Schlottmann}}, \bibinfo{journal}{Phys. Rep.} \textbf{\bibinfo{volume}{181} }\bibinfo{page}{1} (\bibinfo{year}{1989}).
 \end{thebibliography}
\end{document}